\documentstyle[epsf,aps,prl]{revtex} 
\begin{document} 
\draft 

\twocolumn[
\hsize\textwidth\columnwidth\hsize\csname@twocolumnfalse\endcsname

\title{Continuous weak measurement of quantum coherent oscillations } 

\author{A.N. Korotkov and D.V. Averin}

\address{Department of Physics and Astronomy, SUNY Stony Brook, 
Stony Brook, NY 11794-3800} 

\date{\today} 
 
\maketitle 
 
\begin{abstract} 

We consider the problem of continuous quantum measurement of 
coherent oscillations between two quantum states of an 
individual two-state system. It is shown that the interplay 
between the information acquisition and the backaction dephasing 
of the oscillations by the detector imposes a fundamental limit, 
equal to 4, on the signal-to-noise ratio of the measurement. The 
limit is universal, e.g., independent of the coupling strength 
between the detector and system, and results from the tendency 
of quantum measurement to localize the system in one of the 
measured eigenstates. 

\end{abstract} 
\pacs{PACS numbers: 73.23.-b,03.65.Bz}

]
  
Coherent oscillations between the two states of a quantum two-state 
system represent one of the most basic and direct manifestations of 
quantum mechanics and are encountered in practically all areas 
of physics. The question of how to measure them directly in an 
individual two-state system was formulated \cite{b1,b2,b3,b4} for 
the first time in the context of quantum dynamics of Josephson 
junctions, where the oscillating variable, the magnetic flux in a 
superconducting loop, is macroscopic. A common feature of the 
measurement schemes suggested in this context is the use of  
conventional ``projective'' measurements that localize the flux 
in one of its eigenstates and suppress the oscillations. The time 
evolution of the oscillations can then only be studied if the 
experiment is repeated many times with the same initial conditions, 
and the information about oscillations is contained in the 
probability distribution of the 
measurement outcomes. This means that the oscillations are 
effectively studied in an ensemble of systems, not in an individual 
system. Another, practical, disadvantage of the projective 
measurements is the need to switch the detector on and off very 
rapidly, on a time scale shorter than the oscillation period, 
in order to allow for free time evolution of the oscillation and 
subsequent measurement. Since the oscillation frequency is limited 
from below by several factors, including decoherence time and 
temperature, this requirement presents at the very least a challenging 
technical problem. Although this problem can sometimes be solved, 
as demonstrated by the recent observation of coherent quantum 
oscillations of charge \cite{b5}, it is partly responsible for  
the fact that macroscopic quantum oscillations of magnetic 
flux have so far eluded experimental observation. 

The aim of our work is to point out that the problem of 
measurement of quantum coherent oscillations in an individual 
two-state system can be simplified if projective 
measurements are replaced with a weak continuous measurement, and 
to study the quantitative characteristics of such a measurement 
scheme. As is emphasized frequently in the theory of quantum 
measurements (see e.g., \cite{b6,b7,b8}), the ``textbook'' 
projective quantum measurement requires the dynamic interaction 
between the system and the detector to be sufficiently strong to 
establish nearly perfect correlation between their states. If the 
interaction is weak, this does not happen, and the measurement 
provides only limited information about the system. 
Such a weak measurement, however, perturbs the system 
only slightly and can be performed continuously. Below we consider 
quantitatively the process of continuous weak measurement 
of quantum coherent oscillations. We calculate the spectral 
density of the detector output and show that the trade-off between 
the acquisition of information and dephasing due to the detector 
backaction on the oscillations imposes a fundamental limit, equal 
to 4, on the signal-to-noise ratio of the measurement. In this work, 
we use a more conventional non-selective approach to measurement, 
and all the results can be reproduced within the selective 
description of the measurement process \cite{b14}. 

Although the main conclusions of our work are quite general, in 
what follows we prefer to use the language of a particular 
system: two coupled quantum dots measured with a quantum point 
contact. Quantum point contacts were used as electron detectors 
in \cite{b9,b10,br} and described theoretically in 
\cite{b11,b12,b13,b18,b15,b14}. Coherent electron oscillations 
in coupled dots were observed indirectly in the dc transport 
under microwave irradiation \cite{b16}. 

The Hamiltonian of the system (see inset in Fig.\ 1a) is: 
\begin{equation}
H=-\frac{1}{2}(\varepsilon \sigma_z +\Delta \sigma_x +\sigma_z  
U) + \sum_{ik} \varepsilon_k a^{\dagger}_{ik}a_{ik}\, ,
\label{1} \end{equation} 

\vspace{-4ex} 

\[ U=\sum_{ij}U_{ij} \sum_{kp} a^{\dagger}_{ik}a_{jp} \, . \]
The first two terms here describe an electron oscillating between 
the two discrete energy states localized in the quantum dots: 
$\varepsilon$ is the energy difference between the states, 
$-\Delta/2$ is their tunnel coupling, and the $\sigma$'s denote 
Pauli matrices. The 
operators $a_{ik}$ represent point-contact electrons in the two 
scattering states $i=1,2$ (incident from the two contact electrodes) 
with momentum $k$. The coupling $\sigma_z U/2$ is due to an 
additional scattering potential $\pm U(x)/2$ created in the point 
contact by the electron occupying one or the other dot. 
The point contact is biased with a dc voltage $V$, so that changes 
in the scattering potential lead to changes in the current $I$ 
through the contact. We take $eV$ to be much smaller than 
both the Fermi energy in the point contact and the inverse traversal 
time of the contact. This allows us to linearize the energy spectrum 
of the point-contact electrons: $\varepsilon_k= v_F k$, where $v_F$ 
is the Fermi velocity, and neglect the momentum dependence of the 
coupling matrix elements $U_{ij}= \int dx \, \psi_i^*(x) U(x) 
\psi_j(x)$ of the perturbation $U$ in the basis of the two 
scattering states $\psi_i(x)$. We also assume that the $U_{ij}$ are 
sufficiently small for the point contact to operate as a linear 
detector, and treat the contact's response to electron in the dots 
in the linear-response approximation. 

Quantum oscillations of electron between the dots create an 
oscillating component of the current $I$ through the point contact. 
Since the phase of the oscillation diffuses under the backaction 
of the shot noise of the point contact, the oscillations are best 
characterized by their spectral density. To find the spectral 
density of the current $I$ we choose the origin of the coordinate 
$x$ along the contact in such a way that the unperturbed 
scattering potential is effectively symmetric, i.e., the reflection 
amplitudes for both scattering states are the same. Then, the 
current operator calculated at a point $x$ in the asymptotic region 
of the scattering states is: 
\begin{eqnarray} 
I= \frac{e v_F}{L} \sum_{kp} [D(a^{\dagger}_{1k}a_{1p} - 
a^{\dagger}_{2k}a_{2p}) + \nonumber \\ 
i(DR)^{1/2}e^{-i(k-p)|x|} (a^{\dagger}_{1k} 
a_{2p} -a^{\dagger}_{2k} a_{1p}) ] \, ,  
\label{3} \end{eqnarray} 
where $D$ and $R=1-D$ are the transmission and reflection 
probabilities of the point contact, $L$ is a normalization length, 
and the variation of the momentum near the Fermi points (i.e., the 
difference between $k$ and $p$) was neglected everywhere besides the 
phase factor in the second term. The reason for keeping this factor 
will become clear later. 

In the linear-response regime, the current response of the point 
contact is driven by the part of the perturbation $U$ causing 
transitions between the two scattering states $\psi_{1,2}$. 
Considering the effect of this perturbation on the stationary 
(symmetric and antisymmetric) combinations of the scattering states, 
one can show that the real part of the transition matrix 
element $U_{12}$ is related to the change $\delta D$ of the 
transmission probability of the contact: 
\begin{equation} 
U_{12}= \frac{v_F}{L} \frac{\delta D+iu}{(DR)^{1/2}} \, , 
\;\;\;\; U_{21} = U_{12}^*\, . 
\label{4} \end{equation} 
The imaginary part of $U_{12}$, expressed through a dimensionless 
parameter $u$ in eq.\ (\ref{4}), does not affect the current $I$. 
Qualitatively, it characterizes the degree of asymmetry in the 
coupling of the quantum dots to the point contact; $u=0$ if the 
perturbation potential $U(x)$ is applied symmetrically with respect 
to the main scattering potential of the point contact. 

When the point contact is used as a detector in a quantum 
measurement, the current $I$ plays the role of the measurement 
output and should behave classically. This condition requires the 
spectral density of $I$ to be much larger than the spectral density 
of the zero-point fluctuations in the relevant frequency range. It  
is satisfied when the voltage $V$ across the point contact, which 
determines the magnitude and the threshold frequency of the shot 
noise of $I$, is sufficiently large, $eV\gg \varepsilon, \Delta$. 
For the point 
contact to be an effective detector, $eV$ should also be much larger 
than the temperature $T$. In this regime, it is straightforward to 
find the correlation functions of the perturbation $U$ and the 
current $I$ in the zeroth order in $U$ from eqs.\ (\ref{1}), 
(\ref{3}), and (\ref{4}): 
\begin{eqnarray} 
\langle U(t)U(t+\tau) \rangle_0 =\frac{eV}{\pi} \,  
\frac{(\delta D)^2+u^2}{DR} \, \delta(\tau) \, , \label{5} \\  
\langle U(t)I(t+\tau) \rangle_0 = \frac{e^2V}{\pi} 
(i\delta D+u) \, \delta(\tau-\eta)\, . \label{55}
\end{eqnarray} 
The spectral density of $I$ at frequencies much smaller than 
$eV$ is dominated by the regular shot noise, 
and the current correlation function is $K_I^{(0)} (\tau) = 
\langle I(t+\tau)I(t) \rangle_0 =e\langle I \rangle R\delta(\tau)$, 
where $\langle I \rangle= e^2VD/\pi$. The time delay $\eta \equiv 
|x|/v_F$ in eq.\ (\ref{5}) comes from the phase factor 
$e^{-i(k-p)|x|}$ kept in eq.\ (\ref{3}), and is infinitesimally 
small for small traversal time of the contact. It is nevertheless 
important for resolving the ambiguity in averages involving the 
time ordering of $I$ and $U$ that are needed for the calculation 
of the current response: $i\int dt'\langle {\cal T}\{ I(t)U(t') \} 
\rangle_0 = e^2V(\delta D+iu)/\pi$. 

Expression for the current correlation function $K_I(\tau)$ in the 
interaction representation with respect to $U$ is: 
\begin{equation}
 K_I(\tau) = \mbox{Tr} \{ \tilde{\rho} (t) I(t) S^{\dagger}
(t+\tau,t) I(t+\tau) S(t+\tau,t) \} \, , 
\label{7} \end{equation}  
where $\tilde{\rho} (t)$ is the total density matrix of the point 
contact and quantum dots at time $t$, the trace is taken over both 
systems, and $S(t+\tau,t)= {\cal T} \exp \{ (-i/2\hbar) 
\int_t^{t+\tau} dt'\sigma_z U \}$ is the time evolution operator. 
Taking the trace over the electron states in the point contact in 
eq.\ (\ref{7}) with the help of the correlation functions (\ref{5}), 
we get 
\begin{equation} 
K_I(\tau)= K_I^{(0)} (\tau) + \frac{(\delta I)^2}{4} \langle 
\sigma_z \sigma_z(\tau) \rangle \, . 
\label{6} \end{equation} 
The average $\langle\; \cdots \; \rangle $ in eq.\ (\ref{6}) is 
taken over the two states of the quantum dots with the stationary 
dot density matrix $\rho$ established as a result of the interaction 
with the point contact and averaged over its dynamics. The current 
change $\delta I\equiv e^2(\delta D ) V/\pi$ is the current 
response to electron oscillations between the dots, and 
$\sigma_z(\tau)$ now denotes the full time evolution of $\sigma_z$, 
driven both by the dot Hamiltonian and the interaction $U$ with 
the point contact. Qualitatively, eq.\ (\ref{6}) shows that the 
current correlation function directly reflects the correlation 
function of the electron position in the dots given by the 
operator $\sigma_z$. 

The time dependence of the operator $\sigma_z (\tau)$ in eq.\  
(\ref{6}) is obtained by tracing out the point contact degrees of 
freedom in eq.\ (\ref{7}) with the help of the $U$--$U$ correlation 
function (\ref{5}). In this way we get the standard set of equations 
for the matrix elements $\sigma_{ij}$ of $\sigma_z (\tau)$: 
\begin{equation}
\dot{\sigma}_{11}=  \Delta \,\mbox{Im} \sigma_{12}\, , \;\;\;\; 
{\dot\sigma}_{12}=  (i\varepsilon -\Gamma ) \sigma_{12} - i\Delta  
\, \sigma_{11} \, ,  
\label{conv} \end{equation}
and $\sigma_{22}=-\sigma_{11}$. The rate 
\begin{equation} 
\Gamma= eV\frac{(\delta D)^2+u^2}{8\pi DR}
\label{8} \end{equation} 
describes backaction dephasing of the coherent electron oscillations 
between the dots by the point contact. Equation (\ref{8}) shows that 
the dephasing rate reaches a minimum in the case of symmetric 
dot-contact coupling ($u=0$). In this case, the rate of dephasing 
by a point contact has been found in \cite{b11,b12,b13}. Increased 
dephasing in the case of asymmetric dot-contact coupling was 
discussed qualitatively in \cite{b15} and studied experimentally in 
\cite{br}. Since the decrease of $\Gamma$ with decreasing asymmetry 
$u$ does not affect the current response of the point contact, 
symmetric coupling corresponds to an optimum in its operation 
as a detector. In this regime, the point contact represents an ideal 
quantum detector in a sense that the minimum value of the dephasing 
rate (\ref{8}) is determined purely by the rate of information 
acquisition about the state of the quantum dots and can be written 
as $\Gamma=(\delta I)^2/4S_0$, where $S_0=2e\langle I \rangle R$ is 
the spectral density of the current shot noise of the point contact 
\cite{b14}. This part of the dephasing is fundamentally unavoidable 
and reflects the tendency of quantum measurement to localize the 
measured system in one of the eigenstates of the measured observable, 
in our case, the electron position $\sigma_z$.

The dot density matrix $\rho$ satisfies the same set of equations 
(\ref{conv}), except for the normalization, $\rho_{11}+\rho_{22}=1$, 
and its stationary value is $\rho=1/2$. Solving eqs.\ (\ref{conv}) 
with the initial condition $\sigma_z (0)= \sigma_z$ and averaging 
$\sigma_z\sigma_z (\tau)$ over $\rho=1/2$ we find the correlation 
function (\ref{6}) and the spectral density $S_I (\omega ) = 
2 \int_{-\infty}^\infty K_I (\tau )e^{i \omega \tau } d\tau$ for 
$\epsilon=0$: 
\begin{equation}
S_I(\omega ) = S_0 + \frac{\Gamma \Omega^2 (\delta I)^2 }
{(\omega^2-\Omega^2)^2+\Gamma^2\omega^2} \, . 
\label{9} \end{equation}
In the case of biased dots with $\epsilon \neq 0$, it is convenient 
to calculate the spectrum numerically from eq.\ (\ref{conv}). 
The spectrum in this case is plotted in Fig.\ 1 for several values 
of $\epsilon$ and the dephasing rate $\Gamma$. For weak dephasing, 
$\Gamma \ll \Delta$, the spectrum consists of a zero-frequency 
Lorentzian that vanishes at $\epsilon=0$ and grows with increasing 
$|\epsilon |$, and a peak at the oscillation frequency $\Omega= 
(\Delta^2+ \epsilon^2)^{1/2}$. Although the width of the oscillation 
peak is $\Gamma$ and can be small for sufficiently weak dot-contact 
coupling, its height cannot be arbitrarily large in comparison to 
the background noise spectral density $S_0$. At $\epsilon=0$, 
when the amplitude of the oscillations is maximum, the peak height 
is $S_{max}=(\delta I)^2/\Gamma$. Even in this case, the ratio of 
the peak height to the background is limited: 
\begin{equation} 
\frac{S_{max}}{S_0} = \frac{4 (\delta D)^2}{(\delta D)^2+u^2} 
\leq 4 \, . 
\label{10} \end{equation}
This limitation is universal, e.g., independent of the coupling 
strength between the dots and the point contact, and reflects 
quantitatively the interplay between measurement of the quantum 
coherent oscillations and their backaction dephasing. The fact that 
the height of the spectral line of the oscillations can not be much 
larger than the noise background means that, in the time domain, the 
oscillations are drowned in the shot noise. 

\begin{figure}[htb]
\setlength{\unitlength}{1.0in}
\begin{picture}(3.,2.4) 
\put(.0,.0){\epsfxsize=3.in\epsfysize=2.4in\epsfbox{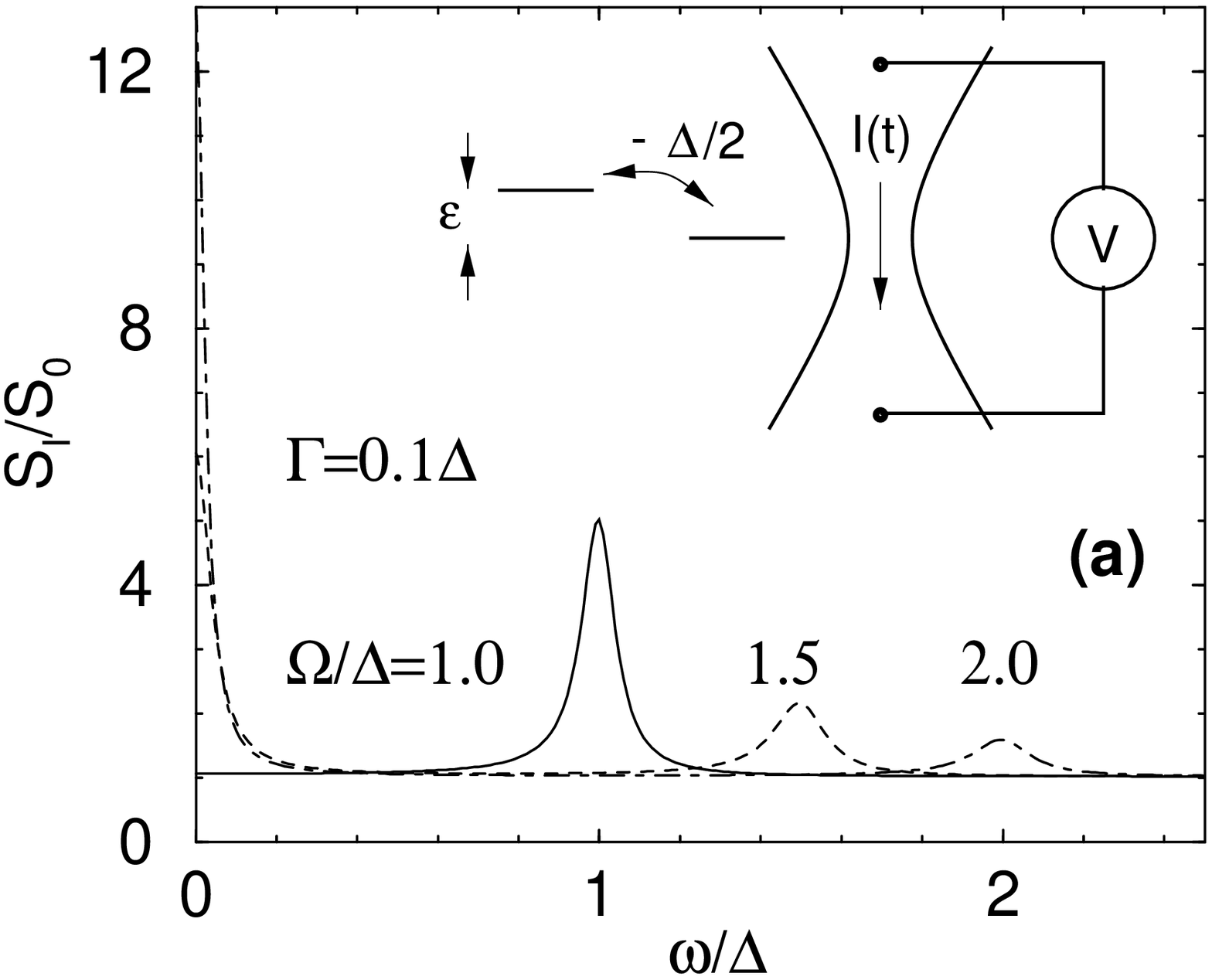}}
\end{picture}

\begin{picture}(3.0,2.4) 
\put(.0,.0){\epsfxsize=3.0in\epsfysize=2.4in\epsfbox{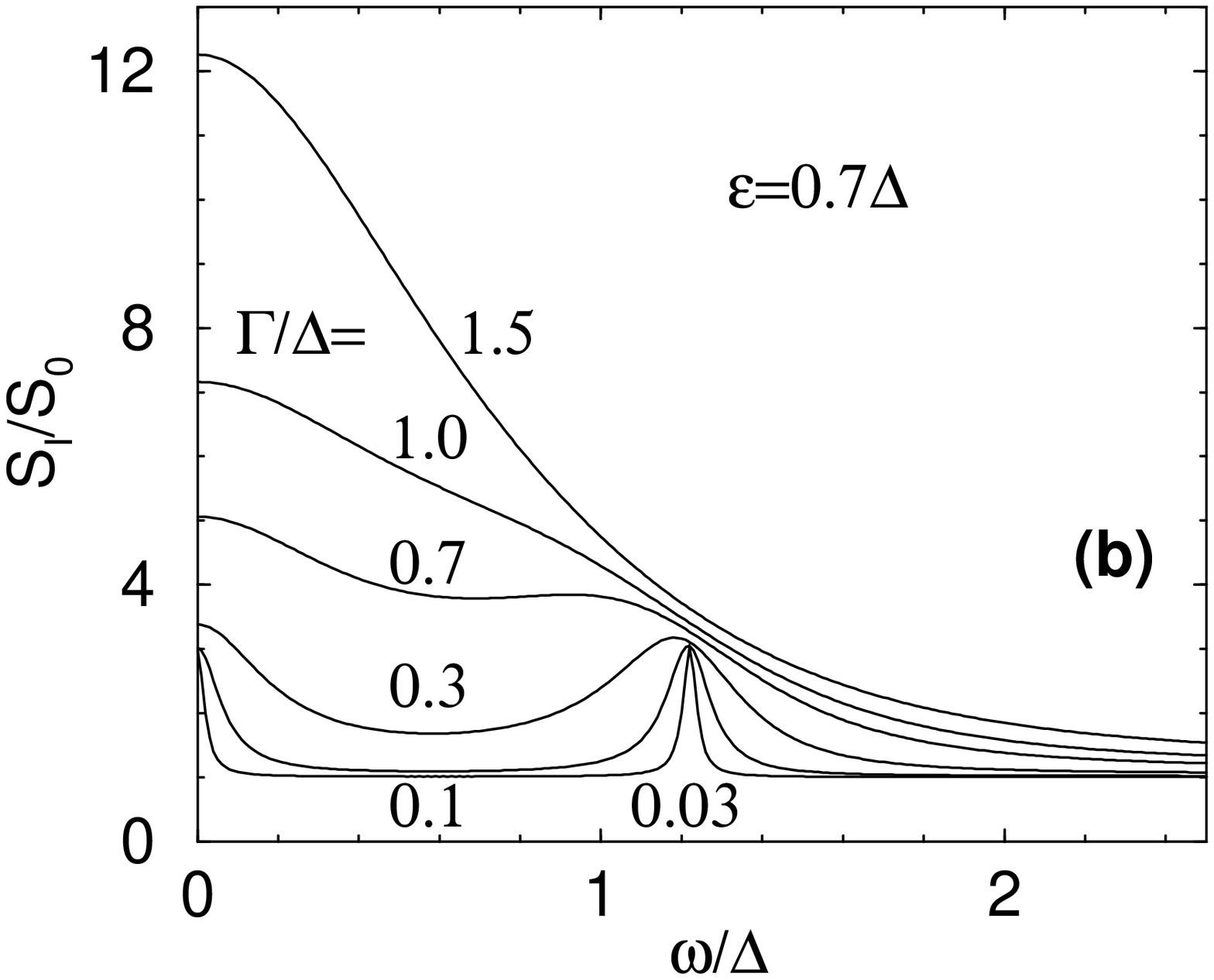}}
\end{picture}

\caption{The diagram (inset in (a)) of the coherent electron 
oscillations between the two discrete energy states in coupled 
quantum dots measured with a point contact. The oscillations are 
detected through modulation of the current $I(t)$ in the point 
contact biased with a voltage $V$. Plotted curves show the spectral 
density $S_I(\omega )$ of the current $I$ in the case of symmetric 
coupling between the point contact and the dots for several values 
of (a) the energy bias $\epsilon$ between the dots reflected in the 
oscillation frequency $\Omega=(\Delta^2+ \epsilon^2)^{1/2}$, and 
(b) the rate $\Gamma$ \protect (\ref{8}) of the measurement-induced 
dephasing. }

\end{figure}

The total intensity of the oscillation line in the spectrum: 
\begin{equation}
\int_0^\infty [S_I(\omega )-S_0] \frac{d\omega}{2\pi} = 
\frac{(\delta I)^2}{4}\, 
\label{11}\end{equation}
does depend on the strength of coupling to the point contact,  
increasing as the coupling becomes stronger. An interesting 
feature of eq.\ (\ref{11}) is that it stresses the impossibility 
of a simple classical interpretation of the quantum coherent 
oscillations, since the intensity of harmonic classical 
oscillations of the same amplitude $\delta I/2$ would be two times 
smaller, and no classical signal of this amplitude could produce 
the oscillation line with intensity (\ref{11}). 

When the backaction dephasing rate $\Gamma$ increases, the 
oscillation line broadens towards the lower frequencies, and 
eventually turns into the growing spectral peak at zero frequency 
that reflects the incoherent electron jumps between the two dots. 
At large $\Gamma$, when the coherent oscillations are suppressed, 
the rate of incoherent tunneling decreases with increasing 
$\Gamma$. For instance, at $\Gamma \gg \Omega$, the tunneling rate 
is $\gamma=\Delta^2/ 
2\Gamma$ , and the spectral density of the point contact response has 
the standard Lorentzian form $S_I(\omega )-S_0=2 \gamma (\delta I)^2
/(4\gamma^2+ \omega^2)$. Suppression of the tunneling rate $\gamma$ 
with increasing dephasing rate $\Gamma$ is an example of the generic 
``Quantum Zeno Effect'' in which quantum measurement suppresses the 
decay rate of a metastable state. In the context of search for the 
macroscopic quantum coherent oscillations, the Lorentzian spectral 
density has been observed and used for measuring the tunneling rate 
of incoherent quantum flux tunneling in SQUIDs \cite{b19}. 
 
The maximum signal-to-noise ratio $S_{max}/S_0$ (\ref{10}) 
is attained if the fundamental backaction of the detector is the only 
dephasing mechanism of the coherent oscillations. In the case of  
measurement with a point contact, the fundamental measurement-induced 
dephasing considered above is created by the backscattering part 
$U_{12}$ (\ref{1}) of the dot--contact interaction that 
dominates at large bias voltages $V$. The forward scattering $U_{11}$, 
$U_{22}$ does not affect the current $I$ in the contact but creates 
a weak additional dephasing and energy-relaxation mechanism for the 
oscillations. We now want to discuss the effect of such a weak 
relaxation on the spectral density of the oscillations noticeable if 
the backaction dephasing is also weak, $\Gamma \ll \Delta$. 

The inclusion of the additional weak relaxation 
does not modify the calculations that lead to eq.\ (\ref{6}), apart 
from a trivial modification that now the average $\sigma_z$ is 
non-vanishing, and the current correlation function should be 
calculated as $K_I(\tau)= K_I^{(0)} (\tau) + (\delta I/2)^2 [(1/2) 
\langle \sigma_z \sigma_z(\tau) +\sigma_z(\tau) \sigma_z\rangle 
- \langle \sigma_z \rangle^2 ]$. For weak coupling, it is convenient 
to find the time evolution of $\sigma_z(\tau)$ in the basis of 
eigenstates of the two-state Hamiltonian $-(\varepsilon \sigma_z + 
\Delta \sigma_x)/2$. Solving the Heisenberg equation of motion up to 
the second order in the dot--contact coupling, and tracing out the 
contact degrees of freedom, we get a set of equations for the 
evolution of the matrix elements $s_{ij}$ of $\sigma_z(\tau)$ in 
the eigenstates basis: 
\[ \dot{s}_{jj}(\tau) = \Gamma_e [\frac{\epsilon}{\Omega} -\coth \{ 
\frac{\Omega}{2T} \} s_{jj}] + (-1)^j \frac{\Gamma \Delta^2}{2\Omega^2} 
(s_{11}-s_{22}) \, , \]

\vspace*{-3ex} 

\begin{equation}
{\dot s}_{12}(\tau) =(i\varepsilon -\Gamma_0 ) s_{12} \, , 
\label{12} \end{equation}
with the initial conditions $s_{11}=-s_{22}=\epsilon/\Omega$, and 
$s_{12}=-\Delta/\Omega$. The characteristic energy-relaxation rate 
in eq. (\ref{12}) is $\Gamma_e =v \Delta^2/\Omega$, where 
$v\equiv (1/\pi) (U_{11}^2+U_{22}^2) (L/v_F)^2$, and the total 
dephasing rate is 
\[ \Gamma_0= [v(\Delta^2 \Omega \coth (\Omega/2T) +4\epsilon^2 T ) 
+ \Gamma (2\epsilon^2+ \Delta^2)]/2\Omega^2 \, .\] 
The dot density matrix $r$ in the eigenstates basis satisfies 
similar equations, and the stationary values of its matrix elements 
are $r_{11}=(\Gamma_e+\Gamma_t)/2\Gamma_t$ and $r_{12}=0$, where 
$\Gamma_t\equiv \Gamma_e \coth (\Omega/2T)+\Gamma \Delta^2/\Omega^2$. 
From these relations and eqs.\ (\ref{12}) we find the spectral 
density: 
\begin{eqnarray}
\lefteqn{S_I(\omega )=S_0 + \frac{(\delta I)^2 }{\Omega^2} \times }
\nonumber \\  
& & \left( \epsilon^2 [1-(\frac{\Gamma_e}{\Gamma_t})^2] \frac{ 
\Gamma_t }{\omega^2+ \Gamma_t^2} + \frac{\Delta^2}{2} \sum_{\pm} 
\frac{\Gamma_0} {(\omega \pm\Omega )^2+\Gamma_0^2 } \right) \, . 
\label{13} \end{eqnarray}
As before, the spectral density consists of a zero-frequency 
Lorentzian and peaks at $\pm \Omega$ of width $\Gamma_0$ that 
represent the coherent electron oscillations. Energy relaxation with 
characteristic rate $\Gamma_e$ broadens the oscillation peak and 
reduces its height $S_{max}$, so that the relative magnitude of the 
peak, $S_{max}/S_0$ decreases in comparison with its value without 
relaxation. 

In summary, we have considered a continuous weak quantum measurement 
by a point contact of quantum coherent oscillations in a two-state 
system, and calculated the spectral density of the output signal of 
the measurement. It has been shown that the backaction dephasing 
introduced into the oscillation dynamics by the measurement imposes 
the fundamental limit on its signal-to-noise ratio. We also 
calculated the effect of energy relaxation on the output spectrum. 

\vspace{1ex} 

The authors are grateful to K.K. Likharev and A.M. van den Brink 
for critical reading of the manuscript. This work was supported in 
part by ARO grant DAAD199910341 and AFOSR grant F496209810025.


\begin{references}
\bibitem{b1} A.J. Leggett and A. Garg, Phys.\ Rev.\ Lett. 
{\bf 54}, 857 (1985). 
\bibitem{b2} L.E. Ballentine, Phys.\ Rev.\ Lett. {\bf 59}, 
1493 (1987)
\bibitem{b3} A. Peres, Phys.\ Rev.\ Lett. {\bf 61}, 2019 (1988). 
\bibitem{b4} C.D. Tesche, Phys.\ Rev.\ Lett. {\bf 64}, 2358 
(1990). 
\bibitem{b5} Y. Nakamura, Yu.A. Pashkin, and J.S. Tsai, Nature 
{\bf 398}, 786 (1999).
\bibitem{b6} Y. Aharonov, D.Z. Albert, and L. Vaidman, Phys.\ 
Rev.\ Lett. {\bf 60}, 1351 (1988).
\bibitem{b7} V.B. Braginsky and F.Ya. Khalili, {\em Quantum 
measurement\,}, (Cambridge University Press, 1992). 
\bibitem{b8} M.B. Mensky, Phys. Usp. {\bf 41}, 923 (1998). 
\bibitem{b14} A.N. Korotkov, Phys.\ Rev. B {\bf 60}, 5737 (1999). 
\bibitem{b9} M. Field, C.G. Smith, M. Pepper, D.A. Ritchie, 
J.E.F. Frost, G.A.C. Jones, and D.G. Hasko, Phys.\ Rev.\ Lett. 
{\bf 70}, 1311 (1993). 
\bibitem{b10} E. Buks, R. Schuster, M. Heiblum, D. Mahalu,
and V.~Umansky, Nature {\bf 391}, 871 (1998).
\bibitem{br} D. Sprinzak, E. Buks, M. Heiblum, and H. Shtrikman, 
cond-mat/9907162. 
\bibitem{b11} S.A. Gurvitz, Phys.\ Rev. B {\bf 56}, 15215 (1997). 
\bibitem{b12} Y. Levinson, Europhys. Lett. {\bf 39}, 299 (1997).
\bibitem{b13} I.L. Aleiner, N.S. Wingreen, and Y. Meir,
Phys.\ Rev.\ Lett. {\bf 79}, 3740 (1997).
\bibitem{b18} G. Hackenbroich, B. Rosenow, and H. A. 
Weidenm\"uller, Phys.\ Rev.\ Lett. {\bf 81}, 5896 (1998). 
\bibitem{b15} L. Stodolsky, Phys.\ Lett. B {\bf 459}, 193 (1999). 
\bibitem{b16} T.H. Oosterkamp, T. Fujisawa, W.G. van der Wiel, 
K.~Ishibashi, R.V. Hijman, S. Tarucha, and L.P. Kouwenhoven, 
Nature {\bf 395}, 873 (1998). 
\bibitem{b19} S. Han, J. Lapointe, and J.E. Lukens,  Phys.\ Rev.\ 
Lett. {\bf 66}, 810 (1991). 

\end{references}
\end{document}